\documentclass[a4paper]{article}
\begin{document}
\title{Two-flux Colliding Plane Waves in String Theory}
\author{Bin Chen\footnote{email:bchen@itp.ac.cn}\\ \\
Interdisciplinary Center of Theoretical Studies, \\
Chinese Academy of Science, P.O. Box 2735\\
Beijing 100080, P.R. China\\
\\
\vspace{3ex}
Jun-Feng Zhang\footnote{email:zhang98@mail.cnu.edu.cn} \\
Department of mathematics, Capital Normal University,\\
Beijing 100037, P.R. China}

 \maketitle
\begin{abstract}
We construct the two-flux colliding plane wave solutions in higher
dimensional gravity theory with dilaton, and two complementary
 fluxes. Two kinds of solutions has been obtained: Bell-Szekeres(BS) type and
 homogeneous type. After imposing the junction condition, we find that
 only Bell-Szekeres type solution is physically well-defined. Furthermore, we show that
 the future curvature singularity is always developed for our solutions.
\end{abstract}

%\tableofcontents
\newpage

\section{Introduction}

The gravitational colliding plane wave (CPW) in four dimensional
gravity has been well studied since its discovery in the early
1970s\cite{cpw,KP}. (For a complete review on four dimensional CPW
solutions, see \cite{Griffiths}) The CPW solutions are the exact
classical solutions of the Einstein gravity, describing the
collision of the plane waves. It is remarkable that a late time
scalar curvature singularity always develop in the CPW solutions,
indicating the fact that the two colliding plane gravitational
wave focus each other so strongly as to produce a spacetime
singularity\cite{Tipler,Yurtsever}. It's true that there exist
infinite families of solutions with a horizon rather than a
curvature singularity. However, such solutions are unstable with
respect to small perturbations in the initial data. Generically, a
curvature singularity forms\cite{Yurtsever}. Therefore, it is
hoped that the study of the CPW solutions may help us to have a
better understanding of the singularity. Also, it has been
proposed that the gravitational plane and dilatonic waves could
play an important role in the pre-big-bang cosmology
scenarios\cite{prebigbang}.

Very recently, there are much interests in the study of CPW
solutions in the higher dimensional gravity. On one side, the
higher dimensional gravity has turned out to have much more
interesting properties than four dimensional gravity. One example
is that the uniqueness and stability issue in higher dimensional
black holes are more complicated\cite{Kodama}.  It has been found
that there exist {\sl black ring} solution with horizon topology
$S^1\times S^2$ in  five dimensional gravity\cite{blackring}. It
could be wished that the CPW solutions in the higher dimensional
gravity have more rich structures and physics. On the other hand,
the plane waves is not only the  classical solutions to vacuum
Einstein equation but also the one to string
theory\cite{planewave}. It is quite interesting to study their
collision in the framework of low energy effective action of
string theory. It is well known that in the low energy effective
action, there are dilaton fields and various kinds of multi-form
fields, coupled with each other in the supergravity action.
Obviously, the complete and thorough discussion of CPW solutions
in higher dimensional gravity is still out of our reach.
Nevertheless, there have been some progress along this direction.
In \cite{Gurses}, Gurses et.al discussed the CPW solutions in
dilaton gravity, in higher dimensional gravity, and in higher
dimensional Einstein-Maxwell theory. In \cite{Pioline}, Gutperle
and Pioline tried to construct the CPW solutions in the
ten-dimensional gravity with self-dual form flux. However, their
generalized BS-type solution (3.37) fails to satisfy the junction
condition and is unphysical. Only in \cite{chen}, the flux-CPW
solution, which named for the CPW solution with flux, has been
successfully constructed in a higher dimensional gravity theory
with dilaton and a higher form flux. Actually, There are two
classes of solutions: one is called $(pqrw)$-type. It looks like
the Bell-Szekeres solutions, which describe the electric-magnetic
CPW in four-dimensional Einstein-Maxwell gravity\cite{BS}. The
other class is a new kind of homogeneous solution, called $(f\pm
g)$ type.

In this paper, we would like to generalize the study in
\cite{chen} to the case with two complementary fluxes in the
theory. We manage to solve the equation of motions, which is more
involved than the ones in \cite{chen} and also find two classes of
solutions generically. But after imposing the junction conditions,
we find that only BS-type solution is physically well defined and
acceptable while the $(f\pm g)$ type solution fails to satisfy the
junction conditions. We also notice that  the future singularity
will always be developed in our solutions.

The paper is organized as follows: in section two, after giving a
brief review on how to find the CPW solution in general, we derive
the equations of motions and solve them; in section three, we
impose the junction condition on the solutions and study the
future singularity of the solution; in section four, we end with
some conclusions and discussions. To be self-consistent, we
include the Riemann and Ricci tensor for our metric ansatz in the
appendix.

\section{Two-flux-CPW solutions to the equations of motions}

In the study of the collision of the gravitational plane waves,
one usually divides the spacetime into four regions: past
P-region($u<0,v<0$), right R-region($u>0, v<0$), left
L-region($u<0, v>0$) and future F-region($u>0,v>0$), which
describes the Minkowski spacetime, the incoming waves from right
and left, and the colliding region respectively. The general
recipe to construct the CPW solutions is to solve the equations of
motions in the forward region and then reduce the solutions to
other regions, requiring the metric to be continuous and
invertible in order to paste the solutions in different regions.
More importantly, one need to impose the junction conditions to
get an  acceptable physical solution. In this section, we focus on
the solutions to the equations of motions and leave the discussion
on the junction conditions to the next section.

We will work in a higher dimensional gravity theory with dilaton
and two complementary fluxes. More precisely, let us consider the
following action with the graviton, the dilaton, a $(n+1)$-form
and a $(m+1)$-form field strength in $D(=n+m+2)$-dimensional
space-time

\begin{equation}\label{action}
S=\int
d^Dx\sqrt{-g}\left(R-g^{\mu\nu}\partial_\mu\phi\partial_\nu\phi-\frac{1}{2(n+1)!}e^{a\phi}F^2
-\frac{1}{2(m+1)!}e^{b\phi}H^2\right)
\end{equation}
where
$F^2=F_{\mu_1...\mu_{n+1}}F^{\mu_1...\mu_{n+1}},H^2=H_{\nu_1...\nu_{m+1}}H^{\nu_1...\nu_{m+1}}$
and $a,b$ are the dilaton coupling constants. Such kind of action
could often appear as the bosonic part of $D>4$ supergravity
theories, which could be the low energy effective actions of
string theory in the Einstein frame or their Kaluza-Klein
reduction. For example, in 5-dim supergravity, one may have
dilaton, gauge 1-form field, and also NS $B_{\mu\nu}$ 2-form
field; in 6-dim supergravity from Kaluza-Klein reduction of 10-dim
supergravity, one may have dilaton, gauge 1-form field and 3-form
field; even in 10-dim supergravity, if m=n=4, our case could be
reduced to the one in Gutperle and Pioline's paper. Usually, the
field content of the $D>4$ supergravity has more fields and
Chern-Simons coupling. After turning off the extra fields and
making the ansatz of the flux field strength to make the
Chern-Simons coupling vanishing, the action could reduce to
(\ref{action}).

From (\ref{action}), the equations of motions are given by
\begin{eqnarray}
R_{\mu\nu}=\partial_\mu\phi\partial_\nu\phi+\frac{1}{2n!}e^{a\phi}\left(F_{\mu\mu_1...\mu_n}F_\nu^
{\ \mu_1...\mu_n}-\frac{n}{(n+1)(m+n)}g_{\mu\nu}F^2\right)+\nonumber\\
+\frac{1}{2m!}e^{b\phi}\left(H_{\mu\nu_1...\nu_m}H_\nu^ {\
\nu_1...\nu_m}-\frac{m}{(m+1)(m+n)}g_{\mu\nu}H^2\right)
\end{eqnarray}
\begin{equation}
\partial_\mu\left(\sqrt{-g}e^{a\phi}F^{\mu\mu_1...\mu_n}\right)=0
\end{equation}
\begin{equation}
\partial_\nu\left(\sqrt{-g}e^{b\phi}H^{\nu\nu_1...\nu_m}\right)=0
\end{equation}
\begin{equation}
\frac{1}{\sqrt{-g}}\partial_\mu\left(\sqrt{-g}g^{\mu\nu}\partial_\nu\phi\right)
=\frac{a}{4(n+1)!}e^{a\phi}F^2+\frac{b}{4(m+1)!}e^{b\phi}H^2
\end{equation}

We make the following CPW ansatz for the metric
\begin{equation}
d^2s=2e^{-M}dudv+e^A\sum_{i=1}^{n}dx_i^2+e^B\sum_{j=1}^{m}dy_j^2,
\end{equation}
and to keep problem tractable, we also restrict ourselves to the
case that the nonzero components of the $(n+1)$-form,$(m+1)$-form
fluxes are
  \begin{eqnarray}
   F_{ux_1...x_n}=C_u \hspace{2cm} F_{vx_1...x_n}=C_v \nonumber\\
   H_{uy_1...y_n}=D_u \hspace{2cm} H_{vy_1...y_n}=D_v
   \end{eqnarray}
 In the sense that the two fluxes $F$ and $H$ occupy mostly the
 $x_i$'s and $y_j$'s respectively, we regard them to be complementary.
We will take the functions $M, A, B, C, D$ and $\phi$ to be function of $u,  v$ only.
The equations of motions for the graviton take
the form
%\footnote{Basically, we follow the notation in \cite{chen}. And the curvature tensor et.al. for the
%metric (6) can be found in the appendix A in \cite{chen}}:

\begin{equation}
nA_{uu}+mB_{uu}+nM_uA_u+mM_uB_u+\frac{1}{2}(nA_u^2+mB_u^2)
=-2\phi_u^2-e^{a\phi-nA}C_u^2-e^{b\phi-mB}D_u^2
\end{equation}
\begin{equation}
nA_{vv}+mB_{vv}+nM_vA_v+mM_vB_v+\frac{1}{2}(nA_v^2+mB_v^2)
=-2\phi_v^2-e^{a\phi-nA}C_v^2-e^{b\phi-mB}D_v^2
\end{equation}
\begin{equation}
-M_{uv}+\frac{n}{2}A_{uv}+\frac{m}{2}B_{uv}+\frac{1}{4}(nA_uA_v+mB_uB_v)
=-\phi_u\phi_v+\frac{n-m}{2(n+m)}e^{a\phi-nA}C_uC_v+\frac{m-n}{2(n+m)}e^{b\phi-mB}D_uD_v
\end{equation}
\begin{equation}
2A_{uv}+nA_uA_v+\frac{m}{2}(A_uB_v+A_vB_u)=-\frac{2m}{n+m}e^{a\phi-nA}C_uC_v
+\frac{2m}{n+m}e^{b\phi-mB}D_uD_v
\end{equation}
\begin{equation}
2B_{uv}+mB_uB_v+\frac{n}{2}(A_uB_v+A_vB_u)=\frac{2n}{n+m}e^{a\phi-nA}C_uC_v
-\frac{2n}{n+m}e^{b\phi-mB}D_uD_v
\end{equation}
 The equations of motions for the dilaton and $n$-form, $m$-form potential
are given by
\begin{equation}
2C_{uv}+\left[a\phi-\frac{1}{2}(nA-mB)\right]_uC_v+\left[a\phi-\frac{1}{2}(nA-mB)\right]_vC_u=0
\end{equation}
\begin{equation}
2D_{uv}+\left[b\phi+\frac{1}{2}(nA-mB)\right]_uD_v+\left[b\phi+\frac{1}{2}(nA-mB)\right]_vD_u=0
\end{equation}
\begin{equation}
 \phi_{uv}+\frac{1}{4}(nA+mB)_u\phi_v+\frac{1}{4}(nA+mB)_v\phi_u
 =\frac{a}{4}e^{a\phi-nA}C_uC_v+\frac{b}{4}e^{b\phi-mB}D_uD_v
\end{equation}

Here we have abbreviated the derivatives by a subscript, e.g.
$A_u=\partial_u A$. As usual, the equation (10) is redundant and will not be needed anymore.
 Introduce
\begin{equation}
U=\frac{1}{2}(nA+mB) \hspace{2cm} V=\frac{1}{2}(nA-mB)
\end{equation}
to make the equations (8), (9), (11), (12) become
\begin{equation}
U_{uu}+M_uU_u+\frac{m+n}{4mn}\left(U_u^2+V_u^2\right)+\frac{m-n}{2mn}U_uV_u
=-\phi_u^2-\frac{1}{2}e^{a\phi-nA}C_u^2
-\frac{1}{2}e^{b\phi-mB}D_u^2
\end{equation}

\begin{equation}
U_{vv}+M_vU_v+\frac{m+n}{4mn}\left(U_v^2+V_v^2\right)+\frac{m-n}{2mn}U_vV_v
=-\phi_v^2-\frac{1}{2}e^{a\phi-nA}C_v^2
-\frac{1}{2}e^{b\phi-mB}D_v^2
\end{equation}
\begin{equation}
U_{uv}+U_uU_v=0
\end{equation}
\begin{equation}
V_{uv}+\frac{1}{2}(U_uV_v+U_vV_u)
=-\frac{mn}{m+n}e^{a\phi-nA}C_uC_v+\frac{mn}{m+n}e^{b\phi-mB}D_uD_v
\end{equation}

Equation (19) is well-known in the study of CPW solutions,  and the general
solution to it is

\begin{equation}
U=\log\left[f(u)+g(v)\right]
\end{equation}
where $f,g$  are arbitrary functions, chosen usually to be
monotonic functions. It is convenient to treat $(f,g)$ as
coordinates alternative to $(u,v)$.

\subsection{Two-flux-CPW solutions to equations of motions: when $a\neq -b$}

In the case that $a \neq -b$, one can define
\begin{equation}
X=a\phi-V \hspace{2cm} Y=b\phi+V
\end{equation}
to simplify  the equations (13), (14) (15) and (20)  in terms of
the $(f,g)$-coordinates. After some linear combinations, one has:

\begin{equation}(f+g)X_{fg}+\frac{1}{2}X_{f}+\frac{1}{2}X_{g}=\frac{1+\delta
a^{2}}{4\delta}e^{X}C_{f}C_{g}-\frac{1-\delta a
b}{4\delta}e^{Y}D_{f}D{g}
\end{equation}
\begin{equation}(f+g)Y_{fg}+\frac{1}{2}Y_{f}+\frac{1}{2}Y_{g}
=-\frac{1-\delta ab}{4\delta}e^{X}C_{f}C_{g}+\frac{1+\delta
b^{2}}{4\delta}e^{Y}D_{f}D_{g}
\end{equation}
\begin{equation}
2C_{fg}+X_{f}C_{g}+X_{g}C_{f}=0
\end{equation}
\begin{equation}
2D_{fg}+Y_{f}D_{g}+Y_{g}D_{f}=0
\end {equation}
 where
\begin{equation}
\delta:=\frac{m+n}{4mn}\leq\frac 12.
\end{equation}
In terms of the $(f,g)$-coordinates, the equations (17) and (18)
can be written as
\begin{equation}
S_{f}+\frac{1}{2}e^{X}C_{f}^{2}+\frac{1}{2}e^{Y}D_{f}^{2}+\frac{(f+g)}{(a+b)^2}\left[(1+\delta
a^2)Y_{f}^{2}+(1+\delta b^{2})X_{f}^{2}+2(1-\delta
ab)X_{f}Y_{f}\right]=0
\end{equation}
\begin{equation}
S_{g}+\frac{1}{2}e^{X}C_{g}^{2}+\frac{1}{2}e^{Y}D_{g}^{2}+\frac{(f+g)}{(a+b)^2}\left[(1+\delta
a^2)Y_{g}^{2}+(1+\delta b^{2})X_{g}^{2}+2(1-\delta
ab)X_{g}Y_{g}\right]=0
\end{equation}
where
\begin{equation}
S=M-(1-\delta)\log(f+g)+\log(f_{u}g_{v})+\eta V
\end{equation}
and
\begin{equation}
\eta:=\frac{m-n}{2mn}
\end{equation}

 The inverse relation of (22) is
\begin{equation}
V=\frac{aY-bX}{a+b}. \hspace{2cm} \phi=\frac{X+Y}{a+b}
\end{equation}

Our strategy here is to solve the above set of coupled
differential equations of $(S, X, Y, C, D)$ as the functions of
$(f, g)$ and then get $(M, A, B, C, D, \phi)$ by straightforward
derivation. As the first step, we need to solve the coupled
differential equations of $(X, Y, C, D)$. Though they seem to be
more complicated than the analogue ones in \cite{chen}, we manage
to find two kinds of solutions with  two different forms of
ansatz.

\begin{itemize}
\item   (pqrw)-type (BS type) solution:\\

   We may try the following ansatz for X,Y,C,D.
\begin{eqnarray}
X=-\log c_{1}\frac{rw+pq}{rw-pq} \hspace{2cm}
Y=-\log c_{2}\frac{rw+pq}{rw-pq}\nonumber\\
C=\gamma_{1}(pw-rq)\hspace{2.5cm} D=\gamma_{2}(pw-rq)
\end{eqnarray}
  where
\begin{equation}
 p:=\sqrt{\frac 12-f} \hspace{1cm}
 q:=\sqrt{\frac 12-g} \hspace{1cm}
 r:=\sqrt{\frac 12+f} \hspace{1cm}
 w:=\sqrt{\frac 12+g}.
\end{equation}

 They satisfy the equations (25) and (26) automatically and from
(23) (24), we have
\begin{equation}
\gamma_{1}^{2}=\frac{8(2+\delta b^{2}-\delta ab)c_{1}}{(a+b)^{2}}
\end{equation}
\begin{equation}
\gamma_{2}^{2}=\frac{8(2+\delta a^{2}-\delta ab)c_{2}}{(a+b)^{2}}
\end{equation}

After integrating (28) and (29) with respect to $f$ and $g$, we
find that
\begin{equation}
S=\frac{4+\delta(a-b)^{2}}{(a+b)^{2}}\left[\log(1-2f)(1+2g)+\log(1+2f)(1-2g)-\log(f+g)\right]
\end{equation}

and then
\begin{equation}
e^{-M}=c_{1}^{\frac{b}{a+b}\eta}c_{2}^{\frac{-a}{a+b}\eta}f_{u}g_{v}
\left[(1-4f^{2})(1-4g^{2})\right]^{-\frac{4+\delta(a-b)^{2}}{(a+b)^2}}
(f+g)^{-\left[1-\delta-\frac{4+\delta(a-b)^{2}}{(a+b)^{2}}\right]}
\left(\frac{rw+pq}{rw-pq}\right) ^{\frac{b-a}{a+b}\eta}
\end{equation}
\begin{equation}
e^{nA}=c_{1}^{\frac{b}{a+b}}c_{2}^{\frac{-a}{a+b}}(f+g)\left(\frac{rw+pq}{rw-pq}\right)^{\frac{b-a}{a+b}}
\end{equation}
\begin{equation}
e^{mB}=c_{1}^{\frac{-b}{a+b}}c_{2}^{\frac{a}{a+b}}(f+g)\left(\frac{rw+pq}{rw-pq}\right)^{\frac{a-b}{a+b}}
\end{equation}
and also the dilaton field is given by
\begin{equation}
e^\phi=\left(c_1c_2\right)^{\frac{-1}{a+b}}\left(\frac{rw+pq}{rw-pq}\right)^{\frac{-2}{a+b}}
\end{equation}

Up to now, we  have solved the equations of motions in the
F-region and find a two-parameter family of solutions depending on
the constants $c_1$ and $c_2$. Actually one can reduce the above
solutions to the ones for the L-region, the R-region, and the
P-region if one do the following replacements:
\begin{equation}
f(u)=f_0\hspace{1cm}f_u(1-2f)^{-\rho}\mid_{f=f_0}=-1\hspace{2cm}
\mbox{for}\
 \ \ u<0
\end{equation}
\begin{equation}
g(v)=g_0\hspace{1cm}g_v(1-2g)^{-\rho}\mid_{g=g_0}=-1\hspace{2cm}
\mbox{for}\
 \ \ v<0
\end{equation}
      where
$\rho=\frac{4+\delta(a-b)^2}{(a+b)^2}$ and
$f_0$, $g_0$ are constants. Taking into account of the continuous and invertible conditions on the metric,
we are able to  fix the values of
\begin{equation}
f_0=g_0=1/2
\end{equation}
and put constraints on the parameters $c_1, c_2$:
\begin{equation}
c_1^b=c_2^a.
\end{equation}
Finally, we only have an one-parameter family of $(pqrw)$-type
solutions.

\item $(f\pm g)$-type solution

   For the second type solution, we take  the
following ansatz for $X ,Y, C$ and $D$  whose dependence in $f, g$  are
of the form $(f\pm g)$:
\begin{eqnarray}
X=X(f+g) \hspace{2cm} Y=Y(f+g)\nonumber\\
C=C(f-g)  \hspace{2cm} D=D(f-g)
\end{eqnarray}
Equation (25)and (26) then gives

\begin{equation}
C=\gamma_1\cdot(f-g) \hspace{2cm}  D=\gamma_2\cdot(f-g)
\end{equation}
for some constants $\gamma_1,\gamma_2$  and (23), (24) can be
solved by
\begin{equation}
X=-\log\left[\frac{\gamma_1^2(a+b)^2}{8c_1^2(2+\delta b^2-\delta
ab)}(f+g)\cosh^2\left(c_1\log{\frac{c_2}{f+g}}\right)\right]
\end{equation}
\begin{equation}
Y=-\log\left[\frac{\gamma_2^2(a+b)^2}{8c_1^2(2+\delta a^2-\delta
ab)}(f+g)\cosh^2\left(c_1\log{\frac{c_2}{f+g}}\right)\right]
\end{equation}
where $c_1, c_2$ are the integration constants. Without loss of
generality one can take $c_1, c_2> 0$. One can then integrate (28)
and (29) to have

\begin{equation}
S=-\frac{4+\delta(a-b)^2}{(a+b)^2}\left[(4c_1^2+1)\log(f+g)+4\log{\cosh\left(c_1\log{\frac{c_2}{f+g}}\right)}
\right]
\end{equation}
and then
\begin{eqnarray}
e^{-M}=\left[\frac{\gamma_1^2(a+b)^2}{8c_1^2(2+\delta b^2-\delta
ab)}\right]^{\frac{b}{a+b}\eta}\left[\frac{\gamma_2^2(a+b)^2}{8c_1^2(2+\delta
a^2-\delta ab)}\right]^{\frac{-a}{a+b}\eta}f_ug_v
(f+g)^{-(1-\delta)+(4c_1^2+1)\frac{4+\delta(a-b)^2}{(a+b)^2}+\frac{b-a}{a+b}\eta}\nonumber\hspace{-2cm}\\
\times\left[\cosh\left(c_1\log{\frac{c_2}{f+g}}\right)\right]
^{\frac{2(b-a)}{a+b}\eta+\frac{4\left[4+\delta(a-b)^2\right]}{(a+b)^2}}
\end{eqnarray}

The other components of the metric are

\begin{equation}
e^{nA}=\left[\frac{\gamma_1^2(a+b)^2}{8c_1^2(2+\delta b^2-\delta
ab)}\right]^{\frac{b}{a+b}}\left[\frac{\gamma_2^2(a+b)^2}{8c_1^2(2+\delta
a^2-\delta ab)}\right]^{\frac{-a}{a+b}} (f+g)^{\frac{2b}{a+b}}
\left[\cosh^2\left(c_1\log{\frac{c_2}{f+g}}\right)\right]
^{\frac{b-a}{a+b}}
\end{equation}
\begin{equation}
e^{mB}=\left[\frac{\gamma_1^2(a+b)^2}{8c_1^2(2+\delta b^2-\delta
ab)}\right]^{\frac{-b}{a+b}}\left[\frac{\gamma_2^2(a+b)^2}{8c_1^2(2+\delta
a^2-\delta ab)}\right]^{\frac{a}{a+b}} (f+g)^{\frac{2a}{a+b}}
\left[\cosh^2\left(c_1\log{\frac{c_2}{f+g}}\right)\right]
^{\frac{a-b}{a+b}}
\end{equation}
and the dilaton field is
\begin{equation}
e^\phi=\left[\frac{\gamma_1^2(a+b)^2}{8c_1^2(2+\delta b^2-\delta
ab)}\right]^{\frac{-1}{a+b}}\left[\frac{\gamma_2^2(a+b)^2}{8c_1^2(2+\delta
a^2-\delta ab)}\right]^{\frac{-1}{a+b}} (f+g)^{\frac{-2}{a+b}}
\left[\cosh^2\left(c_1\log{\frac{c_2}{f+g}}\right)\right]
^{\frac{-2}{a+b}}
\end{equation}

The above solutions give a four-parameters family of solution
labelled by $(\gamma_1, \gamma_2, c_1, c_2)$ in the F-region. In
the same way, they reduce to the solutions in the L-region,
R-region and in the P-region if one do the following replacements

\begin{equation}
f(u)=f_0  \hspace{1.5cm} f_u|_{f_0}=-1 \hspace{2cm}\mbox{for}\ \ u<0
\end{equation}
\begin{equation}
g(v)=g_0  \hspace{1.5cm} g_v|_{g_0}=-1 \hspace{2cm}\mbox{for}\ \
v<0
\end{equation}
A key point here is that the  conditions on $f_u, g_v$ are
different from the ones in $(pqrw)$-type solutions at the
junction. This is due to the fact that at the junction the only
possible singular part  in the metric arises only from $f_u, g_v$
in $e^{-M}$. As we shall see, this fact will restrict the
near-junction expansion of $f$ and $g$ strictly. Similarly, the
patching of the solutions will put constraints on the parameters
$\gamma_1,\gamma_2, c_1, c_2$.

\end{itemize}

\subsection{Two-flux-CPW solutions when $a=-b$}

In the case that $a=-b$, (32) is singular and one needs to change
the variables as follows

\begin{equation}
X=\phi+\delta aV \hspace{2cm} Y=a\phi-V
\end{equation}

Then, in terms of $(f, g)$ coordinates, we have

\begin{equation}
2C_{fg}+Y_{f}C_{g}+Y_{g}C_{f}=0
\end{equation}
\begin{equation}
2D_{fg}-Y_{f}D_{g}-Y_{g}D_{f}=0
\end {equation}
\begin{equation}
(f+g)X_{fg}+\frac{1}{2}(X_f+X_g)=0
\end{equation}
\begin{equation}
(f+g)Y_{fg}+\frac{1}{2}\left(Y_f+Y_g\right) =\frac{1+\delta
a^2}{4\delta}\left(e^YC_fC_g-e^{-Y}D_fD_g\right)
\end{equation}
and also
\begin{equation}
 S_f+\frac{1}{2}e^YC_f^2+\frac{1}{2}e^{-Y}D_f^2+\frac{\delta}{1+a^2\delta}(f+g)Y_f^2+\frac{1}{1+a^2\delta}(f+g)X_f^2=0
 \end{equation}
 \begin{equation}
S_f+\frac{1}{2}e^YC_g^2+\frac{1}{2}e^{-Y}D_g^2+\frac{\delta}{1+a^2\delta}(f+g)Y_g^2+\frac{1}{1+a^2\delta}(f+g)X_g^2=0
\end{equation}
where
\begin{equation}
S=M-(1-\delta)\log(f+g)+\log(f_{u}g_{v})+\eta V
\end{equation}

A special case with $a=-b$ is that $a=b=0$. Then the theory is not
a dilatonic gravity any more and reduces to a gravity theory with
a $n$-form and a $m$-form potential. The CPW solutions of such a
theory haven't been discussed before,  as far as  we know.

Let us first consider the equation on $X$. Note that it takes the same
form as in the standard pure gravitational plane wave collision,
and it can be solved by the Khan-Penrose-Szekeres solution:

\begin{equation}
X=\kappa_1\log\frac{w-p}{w+p}+\kappa_2\log\frac{r-q}{r+q}
\end{equation}
where $\kappa_1$ and $\kappa_2$ are integration constants.

\begin{itemize}
\item $ (pqrw)$-type solution:

We may wish that there exit the same kind of $(pqrw)$-type
solution, even though the set of coupled differential equations on
$(C, D, Y)$ looks more involved than the one flux case.  Our
ansatz are the following:

\begin{equation}
Y=-\log c_1\frac{rw+pq}{rw-pq}
\end{equation}
\begin{equation}
C=\gamma_1(pw-rq)
\end{equation}
\begin{equation}
D=\gamma_2(pw+rq)
\end{equation}
which solves (58) and (59) automatically and from (61) we find
that
\begin{equation}
c_1\gamma_1^2+\frac{\gamma_2^2}{c_1}=\frac{8\delta}{\alpha}
\end{equation}
where
\begin{equation}
\alpha=1+\delta a^2
\end{equation}
and $c_1$ is a constant. After integrating (62) and (63) with X
given by (65), we obtain

\begin{eqnarray}
S=b_1\log(1-2f)(1+2g)+b_2\log(1+2f)(1-2g)+(b_3-1+\delta)\log(f+g)+\nonumber\\
+\frac{2\kappa_1\kappa_2}{\alpha}\log\left(\frac{1}{2}+2fg+2pqrw\right)
\end{eqnarray}
Where
\begin{equation}
b_1=\frac{\kappa_1^2+\delta}{\alpha},\hspace{1cm}
b_2=\frac{\kappa_2^2+\delta}{\alpha},\hspace{1cm}
b_3=1-\delta-\frac{\delta+(\kappa_1+\kappa_2)^2}{\alpha}
\end{equation}

The components of the metric are given by
\begin{eqnarray}
e^{-M}=c_1^\frac{\eta}{\alpha}f_ug_v[(1-2f)(1+2g)]^{-b_1}[(1+2f)(1-2g)]^{-b_2}(f+g)^{-b_3}\times\nonumber\\
\times\left[\frac{1}{2}+2fg+2pqrw\right]^{-\frac{2\kappa_1\kappa_2}{\alpha}}
\left(\frac{rw+pq}{rw-pq}\right)^\frac{\eta}{\alpha}
\left[\left(\frac{w-p}{w+p}\right)^{\kappa_1}\left(\frac{r-q}{r+q}\right)^{\kappa_2}\right]^\frac{a\eta}{\alpha}
\end{eqnarray}

\begin{equation}
e^{nA}=c_1^\frac{1}{\alpha}(f+g)\left(\frac{rw+pq}{rw-pq}\right)^\frac{1}{\alpha}\left[\left(\frac{w-p}{w+p}\right)^{\kappa_1}\left(\frac{r-q}{r+q}\right)^{\kappa_2}\right]^\frac{a}{\alpha}
\end{equation}
\begin{equation}
e^{mB}=c_1^\frac{-1}{\alpha}(f+g)\left(\frac{rw+pq}{rw-pq}\right)^\frac{-1}{\alpha}\left[\left(\frac{w-p}{w+p}\right)^{\kappa_1}\left(\frac{r-q}{r+q}\right)^{\kappa_2}\right]^\frac{-a}{\alpha}
\end{equation}
\begin{equation}
e^\phi=\left(\frac{rw+pq}{rw-pq}\right)^{-\frac{\delta a}{\alpha}}
\left[\left(\frac{w-p}{w+p}\right)^{\kappa_1}\left(\frac{r-q}{r+q}\right)
^{\kappa_2} \right]^\frac{1}{\alpha}
\end{equation}

Actually, taking into account the continuous condition on the
metric, we may fix the value of $c_1=1$ and the above solutions
take exactly the same form as the $(pqrw)$-type flux-CPW solution
in \cite{chen}. So finally we have a three-parameter solutions
labelled by $(\gamma_1(\mbox{or}\gamma_2), \kappa_1, \kappa_2)$.

\item  ($f\pm g$)-type solution:

Here, we may use the same ansatz (45) and have
\begin{equation}
C=\gamma_1\cdot(f-g)\hspace{2cm} D=\gamma_2\cdot(f-g)
\end{equation}
with $\gamma_1,\gamma_2$ being constants. Now the equation on $Y$ reduces to
\begin{equation}
(f+g)Y_{fg}+\frac{1}{2}\left(Y_f+Y_g\right) =-\frac{1+\delta
a^2}{4\delta}\left(\gamma_1^2e^Y-\gamma_2^2e^{-Y}\right).
\end{equation}
Unfortunately, we are not able to solve the above equation
analytically. We will not try to analyze this kind of solution in
the following discussion.

\end{itemize}

\section{Junction conditions and future singularity}

\subsection{Junction conditions on the metric}

The junction conditions play an important role in the discussion
on CPW. Once one has the CPW solutions in the different regions,
one needs to paste these solutions together. Besides the usual
continuous and invertible conditions on the metric, one has to
impose some kind of junction conditions on the metric to get an
acceptable physical solution. More precisely, under the natural
requirement that the stress tensor could be piecewise continuous
instead of being continuous, namely the stress tensor may have
finite jump but not $\delta$-function jump across the junction,
the Ricci tensor is allowed to be piecewise continuous. This leads
to following junction condition on the metric\footnote{For a
detailed discussion on the junction condition, please see
\cite{chen, OS}.}:

 1.  If the metric is $C^1$, then impose the Lichnerowicz condition: the metric has to be at least $C^2$.
Otherwise, if the metric is piecewise $C^1$, then impose the OS
junction conditions\cite{OS} which require :

    \begin{equation}
    g_{\mu\nu},\hspace{1cm}
    \sum_{ij}g^{ij}g_{ij,0},\hspace{1cm}
    \sum_{ij}g^{i0}g_{ij,0},\hspace{2cm}(i,j\neq0).
    \end{equation}
to be continuous across the null surface (note that ``$0$" in the
above formulae stands for $u=0$ or $v=0$). From our ansatz on the
metric, the OS condition means that $U,V,M$ need to be continuous
and $U_u =0$ across the junction at $u=0$. The same happens at the
junction $v=0$.

However, the above Lichnerowicz/OS condition on the metric is not
enough. To be physically sensible, the curvature invariants $R$
and $R_2$ should not have poles at the junction, namely

 2. The curvature invariants $R$, $R_2$ do not blow up at the junction.

Usually, when discussing the CPW solutions, one does not put on
any constraints on $R_{\mu\nu\alpha\beta}, R_4$ or other higher
curvature invariants. We shall not discuss this issue neither.

In the last section, we have constructed the two-flux-CPW
solutions to the equations of motions, we now apply the junction
conditions to these solutions. Before that, let us assume that the
near-junction expansion of $f(u\geq0)$ and $g(v\geq0)$ take the
form:

\begin{equation}
f=f_0(1-d_1u^{n_1})\hspace{2cm}u\sim0^+
\end{equation}
\begin{equation}
g=g_0(1-d_2v^{n_2})\hspace{2cm}v\sim0^+
\end{equation}
The boundary  exponents $n_i$'s indicate the behavior of the
metric near the junction.

\subsection{Imposing  junction  conditions on the $(pqrw)$-type solution}
\begin{itemize}
\item Lichnerowicz/O'Brien-Synge junction conditions\\

In the general case that $a\neq -b$, from the continuous and invertible condition on the metric components we have

\begin{equation}
f_0=g_0 = 1/2.
\end{equation}
\begin{equation}
c_1^{\frac{b}{a+b}\eta}c_2^{\frac{-a}{a+b}\eta}=1\Rightarrow
c_1=c_2^\frac{a}{b}
\end{equation}
from $e^A$ and $e^B$. As for the continuity of $e^{-M}$, the condition (42) requires

\begin{equation}
\rho=1-\frac{1}{n_i},\hspace{2cm}d_i=\left(\frac{2}{n_i}\right)^{n_i},
\hspace{5ex} i=1,2.
\end{equation}

Let us zoom in the behavior of the metric near $u \sim 0, v\sim
0$. Actually it is enough to focus on $u \sim 0$ since the
analysis for $v\sim 0$ is very similar. For $(pqrw)$-type
solution,
 we have for $u\sim0$

\begin{eqnarray}
U_u&=&\left(u^{n_1-1}\frac{-d_1n_1}{1+2g}+l.s.t.\right)\Theta(u)\\
V_u&=&\left(u^{\frac{n_1}{2}-1}e_1(\nu)+l.s.t.\right)\Theta(u)\\
nA_u&=&\left(u^{\frac{n_1}{2}-1}e_1(\nu)+l.s.t.\right)\Theta(u)\\
mB_u&=&\left(-u^{\frac{n_1}{2}-1}e_1(\nu)+l.s.t.\right)\Theta(u)\\
M_u&=&\left(-\eta u^{\frac{n_1}{2}-1}e_1(\nu)+l.s.t.\right)\Theta(u)
\end{eqnarray}
where l.s.t. in the above stands for less singular terms and
$e_0(v),  e_1(v)$ are some nonzero functions of v.  It is easy to see that the metric is $C^1$ if $n_1>2$ and is
piecewise $C^1$ if $n_1 \leq 2$. For the case that metric is $C^1$, the Lichnerowicz condition is
satisfied. As for the case that the metric is piecewise $C^1$,
when $n_1 \leq 2$, we need to impose the  OS
junction conditions, which  require that $U_u$  to be
continuous (i.e. equal to zero) across the junction at u=0.
This leads to the constraint
\begin{equation}
   1 < n_i \leq 2.
   \end{equation}

In the special case that $a=-b$, the continuous and invertible
condition on the metric tell us that
\begin{equation}
c_1=1, \hspace{5ex} \gamma_1^2+\gamma_2^2=\frac{8\delta}{\alpha}
\end{equation}
and
\begin{equation}
b_i=1-\frac{1}{n_i}, \hspace{5ex}
d_i=\left(\frac{2}{n_i}\right)^{n_i}
\end{equation}
where $i=1,2$. With $c_1=1$, we find that the $(pqrw)$-type
two-flux-CPW solutions take the similar from as the one-flux
$(pqrw)$-type solution in \cite{chen}. The analysis of the behavor
near $u\sim 0$ gives the same constraints.

Therefore,  after imposing the Lichnerowicz or the O'Brien-Synge
junction conditions, we have the following allowed possibilities
\begin{equation}
\left\{\begin{array}{l}(1)\ \ 1<n_i\leq 2\hspace{3cm}\mbox{metric
is piecewise $C^1$ }\\(2)\ \ n_i>2\hspace{3.6cm}\mbox{metric is at
least piecewise $C^2$ }\end{array}\right.
\end{equation}
in the $(pqrw)$-type solutions.

\item On $R$,$R_2$\\

As we have discussed, even though the Ricci tensor has no $\delta$-function jump at the junction, one has
to be careful to keep it from blowing up at the junction. Instead of studying the Ricci tensor, equivalently
one can investigate the behavior of the curvature invariants $R, R_2$ at the junctions. Requiring the absence of
blow-up at the junction put more constraints on the $n_i$'s.
From our metric, we know that the scalar curvature has the form
\begin{equation}
R=2e^MR_{uv}+ne^{-A}R_{xx}+me^{-B}R_{yy}
\end{equation}

From the explicit expression of  $R_{uv}, R_{xx}$  and $R_{yy}$, we know that the
 the singularity behavior of R  is controlled by $U_u$,  $V_u,
 M_u$. And $V_u$  is the most singular object which tell us that
 R  is non-singular at the boundary if $n_i\geq 2$

Equivalently,  we can use the equations of motions to find the
following simple form for R
\begin{equation}
R=2e^M\phi_u\phi_v+\frac{m-n}{m+n}\frac{e^{M+X}}{f+g}C_uC_v+\frac{n-m}{m+n}\frac{e^{M+Y}}{f+g}D_uD_v
\end{equation}
The analysis on the boundary behavior  for the field $\phi, C$ and
$D$ gives us the same answer, namely, in order for R not to blow
up at the junction, one requires
that $n_i\geq 2$.\\

The careful discussion on the $R_2$, which is of the form
\begin{equation}
R_2=2e^{2M}R_{uv}^2+2e^{2M}R_{uu}R_{vv}+ne^{-2A}R_{xx}^2+me^{-2B}R_{yy}^2,
\end{equation}
impose the same constraints on $n_i$'s.

\end{itemize}
In summary, after taking into account the constraints from all the
junction conditions, we find the following physical possibilities
:
\begin{eqnarray}
\rho=1-\frac{1}{n_i}, & &\hspace{5ex} \mbox{for}\hspace{3ex} a\neq -b\\
b_i=1-\frac{1}{n_i}, & &\hspace{5ex} \mbox{for}\hspace{3ex} a= -b
\end{eqnarray}
but with the same constraints on the boundary exponents:
\begin{equation}
\left\{\begin{array}{l}(1)\ n_i=2\hspace{3cm}\mbox{metric is
piecewise $C^1$}\\(2)\ n_i>2\hspace{3cm}\mbox{metric is at least
piecewise $C^2$}\end{array}\right.
\end{equation}
on the $(pqrw)$-type solutions.

\subsection{Imposing junction conditions on $(f\pm g)$-type solutions}

Here,  we will only discuss the case when $a\neq -b$. The
continuity of $e^A$ and $e^B$ is automatic. As for the continuity
of $e^{-M}$, the condition (54) requires
\begin{equation}
n_1=1 \hspace{2cm}   d_1=2
\end{equation}

If one fixes the normalization of the metric such that A=B=M=0 in
the P-region, then we get

\begin{equation}
\left[\frac{\gamma_1^2(a+b)^2}{8c_1^2(2+\delta b^2-\delta
ab)}\right]^{\frac{b}{a+b}}\left[\frac{\gamma_2^2(a+b)^2}{8c_1^2(2+\delta
a^2-\delta
ab)}\right]^{\frac{-a}{a+b}}\left[\cosh^2\left(c_1\log{\frac{c_2}{f+g}}\right)\right]
^{\frac{b-a}{a+b}}=1
\end{equation}

\begin{equation}
\left[\frac{\gamma_1^2(a+b)^2}{8c_1^2(2+\delta b^2-\delta
ab)}\right]^{\frac{b}{a+b}\eta}\left[\frac{\gamma_2^2(a+b)^2}{8c_1^2(2+\delta
b^2-\delta
ab)}\right]^{\frac{-a}{a+b}\eta}\left[\cosh^2\left(c_1\log{\frac{c_2}{f+g}}\right)\right]
^{\frac{b-a}{a+b}\eta+\frac{2\left[4+\delta(a-b)^2\right]}{(a+b)^2}}=1
\end{equation}
 One solution of  these constrains is

\begin{equation}
c_2=1
\end{equation}

\begin{equation}
c_1^2=\left[\frac{\gamma_1^2(a+b)^2}{8(2+\delta b^2-\delta
ab)}\right]^{\frac{b}{b-a}}\left[\frac{\gamma_2^2(a+b)^2}{8(2+\delta
a^2-\delta ab)}\right]^{\frac{-a}{b-a}}
\end{equation}

 Similarly, for $u\sim 0$ we have the following expansion
\begin{equation}
U_u=\frac{-2}{1+2g}\Theta(u)
\end{equation}
\begin{equation}
V_u=\left[\frac{2(a-b)}{(a+b)(1+2g)}+\frac{a-b}{a+b}e_1(v)\right]\Theta(u)
\end{equation}
\begin{equation}
nA_u=\left[\frac{-4b}{(a+b)(1+2g)}+\frac{a-b}{a+b}e_1(v)\right]\Theta(u)
\end{equation}
\begin{equation}
mB_u=\left[\frac{4a}{(a+b)(1+2g)}+\frac{a-b}{a+b}e_1(v)\right]\Theta(u)
\end{equation}
\begin{equation}
M_u=\left[\frac{2b_0}{1+2g}-\left(\frac{2\left(4+\delta(a-b)^2\right)}{(a+b)^2}
-\frac{\eta(a-b)}{a+b}\right)e_1(v)\right]\Theta(u)
\end{equation}
Where
\begin{equation}
b_0=-(1-\delta)+(4c_1^2+1)\frac{4+\delta(a-b)^2}{(a+b)^2}-\frac{\eta(a-b)}{a+b}
\end{equation}
\begin{equation}
e_1(v)=2\tanh\left(c_1\log{\frac{2}{1+2g}}\right)\frac{2c_1}{1+2g}
\end{equation}

The key point here is that the continuous condition on the metric requires (55,56), which fix the boundary exponents
completely:
\begin{equation}
n_i=1  \hspace{3cm}i=1,2.
\end{equation}
This means that the metric could only be piecewise $C^1$. The
further imposition of OS condition requires that $U_u$ is
continuous. However, the above expansion $U_u$ is proportional to
the step function, showing the violation of the OS condition.
Therefore, the $(f\pm g)$-type solution is unphysical!

\subsection{Future singularity of the solution}

In  this subsection we will see if the future curvature
singularity will  generically appear in our new higher dimensional
flux-CPW solutions. We will focus on the $(pqrw)$-type solutions,
which are the only physical solutions we have.

First we define a hyper-surface $S_0$:
\begin{equation}
f(u)+g(v)=0
\end{equation}
near which the metric may blow up or vanish. Near $S_0$ we have
\begin{equation}
\frac{rw+pq}{rw-pq}\sim(f+g)^{-1}
\end{equation}
and
\begin{equation}
\frac{w-p}{w+p}\sim (f+g), \hspace{5ex} \frac{r-q}{r+q} \sim (f+g)
\end{equation}

When $a\neq -b$, the singular behavior of the $(pqrw)$-type
solution near $S_0$ read:
\begin{eqnarray}
e^{-M}&\sim &
(f+g)^{-\left[1-\delta-\frac{4+\delta(a-b)^2}{(a+b)^2}-\frac{a-b}{a+b}\eta\right]}\\
e^{nA}&\sim &(f+g)^{\frac{2a}{a+b}}\\
e^{mB}&\sim &(f+g)^{\frac{2b}{a+b}}\\
e^\phi&\sim &(f+g)^\frac{2}{a+b}
\end{eqnarray}

The regularity and invertibility of the metric asks the exponents
on the R.H.S to vanish simultaneously. It is obvious that this is
impossible. Therefore at $f+g=0$ the metric is singular. On the
other hand, the singularity of the metric could be just a
coordinate singularity. To check if the curvature singularity do
appear, let us turn to the curvature invariants $R$, $R_2$ and
$R_4$. Note that, near $S_0$ we have
\begin{equation}
\frac{\partial^{l_1+l_2}M}{\partial{u^{l_1}}\partial{v^{l_2}}}
\sim\frac{\partial^{l_1+l_2}A}{\partial{u^{l_1}}\partial{v^{l_2}}}
\sim\frac{\partial^{l_1+l_2}B}{\partial{u^{l_1}}\partial{v^{l_2}}}
\sim(f+g)^{-(l_1+l_2)}
\end{equation}
Then from the expressions of the Ricci tensor and Riemann tensor
listed in the appendix, it is easy to check that the most singular
terms near $S_0$ in $R^2$ , $R_2$ and $R_4$ are all taking the
following generic form

\begin{equation}
\hspace{3cm}
e^{2M}(f+g)^{-4}\sim(f+g)^{-2\left(1+\delta+\frac{4+\delta(a-b)^2}{(a+b)^2}+\frac{a-b}{a+b}\right)}
\end{equation}

The exponent is
\begin{equation}
-2\left[1+\delta\left(1-\frac{\eta^2}{4\delta^2}\right)
+\delta\left(\frac{a-b)}{a+b}+\frac{\eta}{2\delta}\right)^2+\frac{4}{(a+b)^2}\right]<0
\end{equation}

Therefore the $(pqrw)$-type solutions for $a\neq -b$ will always
develop a late time curvature singularity. So the curvature
singularity will always be developed.

When $a=-b$, the singular behavior of the metric components near
$S_0$ read:
 \begin{eqnarray}
 e^{-M} &\sim &
 (f+g)^{-b_3-\frac{\eta}{\alpha}(1-a(\kappa_1+\kappa_2))}\\
 e^{nA} &\sim & (f+g)^{1-\frac{1}{\alpha}(1-a(\kappa_1+\kappa_2))}\\
 e^{mB} &\sim & (f+g)^{1+\frac{1}{\alpha}(1-a(\kappa_1+\kappa_2))}\\
 e^\phi &\sim & (f+g)^{\frac{1}{\alpha}(\delta
 a+(\kappa_1+\kappa_2))}.
 \end{eqnarray}
Obviously, the exponents above cannot be vanishing at the same
time, indicating the metric is singular at $(f+g)$. And similarly
the singular behavior of $R^2, R_2, R_4$ near $S_0$ is dominated
by
 \begin{equation}
 e^{2M}(f+g)^{-4} \sim
 (f+g)^{2[b_3+\frac{\eta}{\alpha}(1-a(\kappa_1+\kappa_2))-2]}
 \end{equation}
The exponent above could be shown to be less than $-1$, taking
into account of the explicit value of $\eta, \alpha, b_3$ and
$\delta$. This fact shows that the singularity is destined to be
developed in the future.

\section{Conclusions  and  Discussions}

In this paper, we investigate the colliding plane wave solutions
in a higher dimensional dilatonic gravity with two complementary
fluxes, generalize the discussion on flux-CPW solutions in
\cite{chen}. We manage to solve the equations of motions, which
are more complicated than the ones in one flux case. Quite
similarly, using two different ansatz we find
 two types of CPW solutions : $(pqrw)$-type and $(f \pm g)$-type.
The  $(pqrw)$-type solutions  satisfy the
Lichnerowicz/O'Brien-Synge junction conditions and could also keep
the curvature invariants from blowing up at the junction.
Precisely speaking, when $a\neq -b$, we have an one-parameter
family of $(pqrw)$-type solutions, and when $a=-b$ we have a
three-parameter family one.

 Unfortunately, the $(f\pm g)$-type
solutions break the junction conditions and are physically
unacceptable. This fact may indicates that the two-flux background
restrict the system more severe than the one-flux background.
Recall  the set of coupled differential equations on $(X, Y, C,
D)$. Comparing with the corresponding ones in one flux case, it is
obvious that the equations here is more restrictive and harder to
find the solutions. More crucially, in one flux case, the equation
on $X$ takes the same form as the standard one in pure
gravitational colliding plane wave, which is related to Backlund
transformation and inverse scattering method. It has
Khan-Penrose-Szekeres solution whose implications in the metric is
important as emphasized in \cite{chen}. In the two complementary
fluxes case, technically we are short of this kind of solution.
One interesting question is that if we relax the complementary
condition, and take a more general ansatz on the metric, could we
find less restrictive solutions\cite{chen1}?

We have also shown that the physical two-flux $(pqrw)$-type
solutions will always develop a late time curvature singularity,
in consistent with the result in one-flux case. This may reflect
the fact that the strong focus effect of gravity is universal,
even in the higher dimensional gravity theories.

\section*{Acknowledgements:}
 CB is deeply indebted to C.S. Chu, K. Furuta and F.L. Lin for
 many valuable discussions on CPW and their initial interests in
 this project.

\appendix
\section{Riemann and  Ricci  tensors}

In the paper, we make the following ansatz to the metric
\begin{equation}
ds^2=2e^{-M}dudv+e^A\sum_{i=1}^ndx_i^2+e^B\sum_{j-1}^mdy_j^2
\end{equation}
with the functions $M, A,  B$  being functions of $u,  v$. We have
Ricci tensor

\begin{eqnarray}
R_{uu}&=&-\frac{1}{2}\left[nA_{uu}+mB_{uu}+nM_uA_u+mM_uB_u+\frac{1}{2}(nA_u^2+mB_u^2)\right]\\
R_{vv}&=&-\frac{1}{2}\left[nA_{vv}+mB_{vv}+nM_vA_v+mM_vB_v+\frac{1}{2}(nA_v^2+mB_v^2)\right]\\
R_{uv}&=&M_{uv}-\frac{n}{2}A_{uv}-\frac{m}{2}B_{uv}-\frac{1}{4}(nA_uA_v+mB_uB_v)\\
R_{xx}&=&-\frac{1}{2}e^{M+A}\left[2A_{uv}+nA_uA_v+\frac{m}{2}(A_uB_v+A_vB_u)\right]\\
R_{yy}&=&-\frac{1}{2}e^{M+B}\left[2B_{uv}+nB_uB_v+\frac{n}{2}(A_uB_v+A_vB_u)\right]
\end{eqnarray}
where $x = x_i$ with $i = 1, \cdots, n$ and $y = y_j$ with $j = 1,
\cdots,m$. And also we have the independent non-vanishing
components of the Riemann tensor as following:

\begin{eqnarray}
R_{uvuv}&=&-e^MM_{uv}\\
R_{xyxy}&=&-\frac{1}{4}e^{M+A+B}(A_uB_v+A_vB_u)\\
R_{uxvx}&=&-e^A(\frac{1}{2}A_{uv}+\frac{1}{4}A_uA_v)\\
R_{vxvx}&=&-e^A(\frac{1}{2}A_{vv}+\frac{1}{2}M_vA_v+\frac{1}{4}A_v^2)\\
R_{uxux}&=&-e^A(\frac{1}{2}A_{uu}+\frac{1}{2}M_uA_u+\frac{1}{4}A_u^2)\\
R_{uyvy}&=&-e^B(\frac{1}{2}B_{uv}+\frac{1}{4}B_uB_v)\\
R_{vyvy}&=&-e^B(\frac{1}{2}B_{vv}+\frac{1}{2}M_vB_v+\frac{1}{4}B_v^2)\\
R_{uyuy}&=&-e^B(\frac{1}{2}B_{uu}+\frac{1}{2}M_uB_u+\frac{1}{4}B_u^2).
\end{eqnarray}

\end{document}